# Dynamically Tie the Right Offer to the Right Customer in Telecommunications Industry

By Kunal Sawarkar, IBM India Software Lab - SWG, Business Analytics, India

Sanket Jain, GBS Business Analytics and Optimization Center of Competence, CMS Analytics India

## ABSTRACT

For a successful business, engaging in an effective campaign is a key task for marketers. Most previous studies used various mathematical models to segment customers without considering the correlation between customer segmentation and a campaign. This work presents a conceptual model by studying the significant campaign-dependent variables of customer targeting in customer segmentation context. In this way, the processes of customer segmentation and targeting thus can be linked and solved together. The outcomes of customer segmentation of this study could be more meaningful and relevant for marketers. This investigation applies a customer life time value (LTV) model to assess the fitness between targeted customer groups and marketing strategies. To integrate customer segmentation and customer targeting, this work uses the genetic algorithm (GA) to determine the optimized marketing strategy. Later, we suggest using C&RT (Classification and Regression Tree) in SPSS PASW Modeler as the replacement to Genetic Algorithm technique to accomplish these results. We also suggest using LOSSYCOUNTING and Counting Bloom Filter to dynamically design the right and up-to-date offer to the right customer.

Search keywords: Genetic Algorithm; C&RT; SPSS; LOSSYCOUNTING; Counting Bloom Filter.

## 1. INTRODUCTION

For a successful business, engaging in an effective campaign is a key task for marketers. Traditionally, marketers must first identify market segmentation using a mathematical model and then implement an efficient campaign plan to target profitable customers (Fraley & Thearting, 1999). This process confronts considerable problems. First, most previous studies used various mathematical models to segment customers without considering the correlation between customer segmentation and a campaign. Previously, the link between customer segmentation and campaign activities was most manual or missing (Fraley & Thearting, 1999). For marketing researchers, segmentation should not be the end in itself, but rather a means to an end (Jonker, Piersma, & Poel, 2004). Following the notion proposed by Jonker, this work presents a conceptual model by counting the significant campaign dependent variables of customer targeting in customer segmentation. In this way, the processes of customer segmentation and targeting thus can be linked and solved together. The outcomes of customer segmentation of this study could be more meaningful and relevant for marketers. To solve the core problem of marketers facing, this investigation applies a customer life time value (LTV) model to assess the fitness between targeted customer groups and marketing strategies. To integrate customer segmentation and customer targeting, this work uses the genetic algorithm (GA) to determine the optimized marketing strategy (Jonker et al., 2004; Kim & Street, 2004; Kim et al., 2005; Tsai & Chiu, 2004). We can explore using C&RT (Classification and Regression Tree) in SPSS PASW Modeler as a substitute for genetic algorithm to accomplish this objective.

## 2. BACKGROUND AND RESEARCH MOTIVATION

According to Chan (2008), most previous studies classified RFM and LTV models as two different methods of segmenting customers. Generally, RFM models represent customer dynamic behavior. On the other hand, LTV models normally evaluate customer value or contribution. Identifying the behavior of high-value customers is a key task in customer segmentation research. This study proposes an intelligent model that uses GA to select customer RFM behavior using a LTV evaluation model. Then it explores the possibility of C&RT model in SPSS as a potential replacement. Customer life time value is taken as the fitness value of GA. If the proposed methodology is applied, high-value customers can be identified for campaign programs. Another advantage of the proposed methodology is that it considers the correlation between customer values and campaigns. Valuable customers thus can be identified for a campaign program.

However, this approach of using genetic algorithm has some limitations. First, this proposed method requires numerous customer data. Second, if you need more breakpoints for your variables, then it can quickly become a very complex exercise. In the future, more breakpoints can be investigated to determine the optimal numbers of breakpoints for customer segmentation. Finally, future studies should also explore the possibility of targeting and finding new customers.

This work will try to focus on devising an approach for dealing with customer segmentation problems during a promotion campaign for an existing customer base.

Currently, in telecom, we typically design only a handful of offers for the entire customer base. These offers can be for retention, revenue maximization, or upgrade. But it would be worth exploring if we can tailor-make offers for each customer, or at least increase the number of offers so that more customers can be fitted to their offer. However, the current scenario does not allow you to map the offer to your individual customer's usage. To do this, we need to adopt the following methodology.

Take as input the customer demographic data along with profile data that includes certain keywords taken from his e-billing statement. To draw a parallel with the credit card and banking industry, let us see how it can be accomplished, and then we can design a similar approach for telecommunications industry.

Scenario: User logs onto ICICI bank portal to check his mini-statement. There he finds that a salary of $100,000 has been credited to his account this month. He might also find that $30,000 home loan EMI had been deducted, and that $5,000 car loan got deducted too. So, in real time, if the application can calculate his net savings (say, $10,000), and then dynamically offer him a tailor-made credit card that would encourage him to spend in the region of $10,000.

*NOTE: The keywords are underscored.*

Along with this, all the banking corporations keep a standard keyword dictionary of terms like ATM withdrawal codes, EMI, car loan, etc. So, they can easily map the keyword searched to its corresponding term in their standard keyword dictionary of terms.

*Note: The specific illustration of offering a credit card based on his real-time income vs. expenses assessment. Further, this should also leverage the entire 360 degree view of the customer to include his average credit card expenses (if already has a card with same bank), payment history and ratings (bureau data).*

## 3. APPROACH

According to Chan (2008), first the companies must plan and develop a marketing strategy for promotion campaigns this year for the existing customer base. Second, marketers must gather customer data to establish customer profile and devise associated campaign information. This study uses a RFM model to represent customer behavior. Customer data are encoded using a RFM model and transformed into a binary string as the input format of genetic algorithm (GA). Conversely, this study collects and calculates customer lifetime values as the fitness values of GA. The proposed LTV model considers the correlation between campaign strategy and customer value. The fourth step is segmenting customers into several homogenous groups using GA. Te fifth step involves targeting and matching segmented customers with the developed campaign strategies and programs. The final step involves classifying customers and turning a campaign plan into action.

Apart from this, it could be interesting to explore whether there exist clearly distinct categories of customers according to their status of offer. For instance, there could be some customers (A) who have never qualified for any offer in the last six months. Then some (B) have qualified for some offer(s) in the last six months but have either rejected or not accepted it. Remaining (C) have accepted their offer(s). So, using this Genetic Algorithm or C&RT, you might further need to customize your offers according to these segments.

Also, it would be better if the system can be trained with the previous campaign feedback. For example, a customer who has never qualified for any offer in the past six months should be treated differently from the one who has accepted all the offers that have been sent to him in the same timeframe.

Now, after using the RFM model to represent customer behavior, Chan (2008) encodes data into a binary string by dividing the values of Recency (R), Frequency (F) and Monetary value (M) into five sections, 0-20%, 20-40%, 40-60%, 60-80% and 80-100%. If the value lies between 20% and 40% the binary code is set to 2. Similarly, if the value is between 60% and 80% the binary code is 4. By doing this, a mapping can be established between the input data and binary codes. This proposed encoding scheme transforms the points of the parameter space into a binary string representation. For instance, a point (1, 2, 5) in a three-dimensional parameter space can be represented as a concatenated binary string (Jang, Sun, & Mizutani, 1997).

0001 0010 0101 can be translated using genetic algorithm to 1 2 5.

*Note: The bits have been underscored only for illustrative purposes.*

At first, the input parameters must be set up and customer data must be encoded as a binary string. Second, GA initializes the chromosome randomly. Third, each chromosome is evaluated. Fourth, higher fitness value members are selected as parents for the next generation. Fifth, crossover is used to generate new chromosomes with provable crossover rate that we hope to preserve good genes from parents. Sixth, mutation is used to flip a bit with the probability of fixed mutation rate. This step can generate new chromosomes to prevent the entire population from converging on trapped local optima. Seventh, a new generation is produced. Meanwhile, the eighth step is evaluating a new generation to measure the stop criteria. If the stop criteria remain unsatisfied, the processes will repeat. If the criteria are satisfied, the evolution will stop. Finally, the best chromosomes are chosen and decoded as the final solutions. The method of segmentation in this study is defined by variable breakpoints (Jonker et al., 2004). The number of segments increases rapidly whenever the number of breakpoints increases.

Later in this study, we need to explore how we can use C&RT algorithm of SPSS to achieve similar results.

To mirror the process of genetic algorithm using C&RT approach, all the relevant input predictor variables would need to be encoded akin to the Genetic Algorithm's 0-1 format. For doing this, we will have to transform each (continuous) predictor variable first into an ordinal variable, and then into multiple binary variables. At the same time, if there is a predictor variable having only two possible values (such as the variable Gender could have value Male or Female), then those values (in this case of Male and Female) can be directly transformed to 1 and 0, respectively.

**OVERVIEW OF C&RT**

C&RT modeling is an exploratory data analysis method used to study the relationships between a dependent measure and a large series of possible predictor variables with potential interactions among themselves. The dependent measure may be a qualitative (nominal or ordinal) one or a quantitative indicator. We can use a Gini measure of association, and prune the branches.

## READING THE OUTPUT DIAGRAM OF C&RT

In C&RT diagram, a series of predictor variables are assessed to see if splitting the sample based on these predictors leads to better discrimination in the dependent measure. For instance, if our dependent measure is whether the patient has gotten medical case management services, we would first assess whether there are different levels of receiving this service for two groups formed on the basis of one of the predictor variables. The most significant of these predictions would define the first split of the sample, or the first branching of the tree. Then, for each of the new groups formed, we would ask if the subgroup could be further significantly split by another of the predictor variables. This process will continue. After each split, we ask if the new subgroup can be further split on another variable so that there are significant differences in the dependent variable. The result at the end of the tree building process is that we have a series of groups that are maximally different from one another on the dependent variable. At each step, the optimal binary split is made. Different orientations of the same tree are sometimes useful to highlight different portions of the results. Here, all splits are binary. However, the same variable may be split repeatedly.

C&RT method has certain advantages as a way of looking for patterns in complicated datasets. First, the level of measurement for the dependent variable and predictor variables can be nominal or ordinal (categorical) or interval (a "scale"). Also, the missing values in predictor variables can be estimated from other predictor ("surrogate") variables, so that partial data can be used whenever possible within the tree. But, C&RT modeling is essentially a "stepwise" statistical method, and there is always a potential for too much to be seen in the data even when very conservative statistical criteria are used.

For us, the dependent variable could be whether the offer is relevant for a customer or not.

*Note: There is no guarantee that C&RT model will yield good accuracy.*

## 4.  DATA AND VARIABLES

Now, taking this analogy to **telecom**, consider the case of a customer who opts for a Vodafone recharge of his prepaid account. So, the application will offer him an upgrade by offering him a postpaid connection. So, in a way, the application system will automatically find business rules.

For e.g., if there is a customer who always spends $10 only for taking a broadband connection, then a business rule should automatically get created for him. An example of offer could be: "unlimited access to Facebook by spending $10 upfront".

To do all this, we will have to keep an account of the billing statement of the customer's mobile phone. This would be akin to his bank mini-statement. Some keywords in a POST-PAID mobile phone connection's statement could be:
   1.  Voice calls - outgoing local to Airtel mobile, to other mobiles, to fixed landline,

2. Voice calls - outgoing STD to Airtel mobile, to other mobiles, to fixed landline,
3. Voice calls – outgoing <u>ISD</u>,
4. Last bill period <u>late fee</u>,
5. Value Added Services: <u>SMS</u> – local to other mobiles,
6. Value Added Services: SMS – <u>national</u> to <u>Airtel</u> mobiles,
7. CUG (this indicates the social network of your customer), and
8. XXX nat sms free (this is "National SMS free" <u>discount package</u>), where XXX could be 100, 150, 200, etc.

*Note: The important keywords have been underscored.*

Similar analysis can be conducted for pre-paid connection.

Also, the next question would arise: How real-time can we make this. To my understanding, I think we can do batch processing because we are not being time-specific or location-bound here, such as offering dinner coupon or late evening movie ticket at Bangalore if you are currently at Bangalore.

Actually, you might not need any bills or statements at all. All the data that is used to generate the statements is anyways present in the DW (data warehouse). So, the analysis can be done directly with the data from the DW (TDW in this case for IBM).

Key questions worth exploring would be as follows:
- How does data analytics drive my product/offer/solution innovation?
- How does an analytics-based rule link with the right product/service/campaign offer mix?
- How can analytics get closer to the product or marketing mix development activity?

Here is another example: on your Gmail account, when you are looking for GMAT website (such as www.gmat.com), it shows you the links to Princeton Review and GRE and other relevant information. Can we do the same for Telecom?

Here is a list of offers that can be given to customers (after having mapped his profile with your strategy or campaign objective):

1. Offer free talk time of say, 50 minutes (if no free time exists as of now).
2. Or, paid talk time of say, 100 minutes by paying $10 upfront (it he is a habitual $10 payer), if there exist enough relevant keywords in his billing statement.
3. VAS offer.
4. Discount package – based on 150 nat sms free (i.e., 150 national SMS free).
5. Upgrade offer (if no late fee in the last 6 months).

The challenge here is that unlike in banking industry, there is no common database of standardized terms that can be used to map these keywords. CDR (Call Detail Record) is our best bet.

But, deploying or even understanding a solution based on genetic algorithm can become quite complex. So, we can use LOSSYCOUNTING to find most frequent business rules. We start with knowing the campaign objective (e.g. retention) and then its parameters (e.g., churn probability, days since last upgrade, days of zero usage). Based on these parameters, the Response Log feature of Telecom Packaging in Decision Management will come up with certain number of business rules. We can then find the most effective (or, most 'popular') business rules by looking for the most frequently occurring business rule using LOSSYCOUNTING. Next, if a rule is "age < 27 *and* margin amount > $30", then it should be *promoted* and other inferior rule like 'age < 27 *and* margin amount < $10" should be *retired*. This can be accomplished by LOSSYCOUNTING. In real life, there could be many business rules, hence selecting the most effective ones is an objective that LOSSYCOUNTING can help us attain. Example offers could be:

- If the most effective business rule is "age < 27 years *and* margin amount > $30 *and* heavy night SMS user", then offer him 100 Free night SMS.
- If the most effective business rule is "age < 24 years *and* margin amount < $10 *and* outgoing call minutes < 25 *and* heavy Internet user", then offer her one month access to Facebook.com for $10 upfront fee. (Notice how his low usage was 'mapped' to give an offer for a $10 upfront fee).

Let us see how the LOSSYCOUNTING algorithm works:

Imagine that you see a large number of individual transactions (such as Amazon book sales), and you want to calculate what are the top sellers today. Or imagine that you are monitoring network traffic and you want to know which hosts/subnets are responsible for most of the traffic. This is a problem of finding *heavy hitters* given a stream of elements. An easy method to solve this problem is to store each element identifier with a corresponding counter monitoring the number of occurrences of that element. Then, you sort the elements accordingly to their counters and you can easily get the most frequent elements. However, in many real scenarios, this simple solution is not efficient of computationally feasible. For instance, consider the case of tracking the pairs of IP address that generate the most traffic over some time period. You need 16,384 PBytes of memory and a lot of time to sort and scan that memory array, which often makes the problem non-computationally feasible. Hence, during recent years, techniques to computing *heavy hitters* using limited memory resources have been investigated. They cannot find exact *heavy hitters*, but instead approximate the *heavy hitters* of a data stream. The approximation typically lies in that the computed *heavy hitters* may include false negatives. LOSSYCOUNTING is an algorithm for finding heavy hitters using limited memory. Here, an important parameter for each distinct element identifier in the table is its *error bound*. Such error bound reflects the potential error on the estimated frequency of an element. Elements with small error bounds are more likely to be removed from LOSSYCOUNTING process than equal-frequency elements having a larger error bound.

The LOSSYCOUNTING algorithm was proposed by Manku and Motwani in 2002, in addition to a randomized sampling-based algorithm and techniques for extending from frequent items to frequent item-sets. The algorithm stores tuples which comprise an item, a lower bound on its count, and a 'delta' value which records the difference between the upper bound and the lower bound. When processing the ith item, if it is currently stored then its lower bound is increased by one; else, a new tuple is created with the lower bound set to one, and _delta set to [i/k]. Periodically, all tuples whose upper bound is less than [i/k] are deleted. These are correct upper and lower bounds on the count of each item, so at the end of the stream, all items whose count exceeds n/k must be stored. As with FREQUENT algorithm, setting k = 1/epsilon ensures that the error in any approximate count is at most en, where e = epsilon. A careful argument demonstrates that the worst case space used by this algorithm is O(1/elogen), and for certain input distributions it is O(1/e).

Now that we know how LOSSYCOUNTING works, we should ensure that we tie that with a mechanism that inserts and deletes rules.

Hash function can be used to lookup for most commonly occurring. However, it may map two or more keys to the same hash value. We wish to minimize the occurrence of such collisions; hence the hash function must map the keys to the hash values as evenly as possible. For any choice of hash unction, there exist bad set of hash keys that all hash to the same slot. The idea is to use hash function at random, independent from the keys. (Remember that the probability of any two keys from colliding each other is less than 1/2).

When testing a hash function, the uniformity of the distribution of hash values can be evaluated by chi-square test. Bloom filter is a probabilistic data structure that uses the hash concept to test whether an element is a member of a set. An empty Bloom filter is a bit array of $m$ bits, all set to 0. There must also be $k$ different hash functions defined, each of which maps or hashes some set element to one of the $m$ array positions with a uniform random distribution. To add an element, feed it to each of the $k$ hash functions to get $k$ array positions. Set the bits at all these positions to 1. To query for an element (test whether it is in the set), feed it to each of the $k$ hash functions to get $k$ array positions. If any of the bits at these positions are 0, the element is not in the set – if it were, then all the bits would have been set to 1 when it was inserted. If all are 1, then either the element is in the set, or the bits have been set to 1 during the insertion of other elements. Removing an element is impossible.

So, we will use Counting Bloom Filter (CBF) to perform insertion & deletion of business rules. This will automatically and dynamically *promote* the best rules and *retire* the others. CBF provides a way to implement a *delete* operation on a Bloom filter without recreating the filter afresh. In a counting filter the array positions (buckets) are extended from being a single bit, to an n-bit counter. The insert operation is extended to *increment* the value of the buckets and the lookup operation checks that each of the required buckets is non-zero. The delete operation, obviously, then consists of decrementing the value of each of the respective buckets.

The advantage of using a Bloom Filter is that while risking false positives, it has a strong space advantage over other data structures for representing sets, because it does not require storing data. Mining frequent *item-sets* inherently builds on finding frequent items as a basic building block. In our case, we intend to find the most frequent business rules and/or the most frequent keywords within those rules. Finding the entropy of a stream requires learning the most frequent items in order to directly compute their contribution to the entropy, and remove their contribution before approximating the entropy of the residual stream.

So, we need to promote the most frequent (and hence, the most effective) business rules and retire the ones that are not frequent. Our technique will use hashing (CBF) to derive multiple sub-streams, the frequent elements of which will be extracted to estimate the frequency moments of the stream using LOSSYCOUNTING.